\newcounter{myctr}
\def\myitem{\refstepcounter{myctr}\bibfont\noindent\ifnum\themyctr>9\else\phantom{0}\fi\hangindent17pt\themyctr.\enskip}

\documentclass{ws-ijqi}
\usepackage{graphicx}

\newcommand{\ket}[1]{|#1\rangle}
\usepackage{color}

\begin{document}

\markboth{B. Bellomo, G. Compagno, R. Lo Franco, A. Ridolfo and S. Savasta}
{Dynamics and extraction of quantum discord in a multipartite open system}

\catchline{}{}{}{}{}

\title{DYNAMICS AND EXTRACTION OF QUANTUM DISCORD IN A MULTIPARTITE OPEN SYSTEM}

\author{B. BELLOMO$^1$, G. COMPAGNO$^1$, R. LO FRANCO$^{1,2}$, A. RIDOLFO$^3$ AND S. SAVASTA$^3$}

\address{$^1$Dipartimento di Fisica, Universit\`a di Palermo, via Archirafi 36, 90123 Palermo, Italy.\\
bruno.bellomo@fisica.unipa.it}

\address{$^2$Centro Siciliano di Fisica Nucleare e di Struttura della Materia (CSFNSM) and Dipartimento di Fisica e Astronomia, Universit\`a di Catania, Viale A. Doria 6, 95125 Catania, Italy}

\address{$^3$Dipartimento di Fisica della Materia e Ingegneria Elettronica, Universit\`{a} di Messina Salita Sperone 31, I-98166 Messina, Italy}

\maketitle

\begin{history}
\received{Day Month Year}
\revised{Day Month Year}
\end{history}

\begin{abstract}
We consider a multipartite system consisting of two noninteracting qubits each embedded in a single-mode leaky cavity, in turn connected to an external bosonic reservoir. Initially, we take the two qubits in an entangled state while the cavities and the reservoirs have zero photons. We investigate, in this six-partite quantum system, the transfer of quantum discord from the qubits to the cavities and reservoirs. We show that this transfer occurs also when the cavities are not entangled. Moreover, we discuss how quantum discord can be extracted from the cavities and transferred to distant systems by traveling leaking photons, using the input-output theory.
\end{abstract}


\section{Introduction}

In the study of correlations present in quantum states two different approaches are nowadays typically distinguished: one is based on the entanglement-versus-separability paradigm and the other on a quantum-classical dichotomy.\cite{Luo2008bPRA}
In the entanglement-separability paradigm introduced by Werner\cite{werner1989PRA}, the state of a multipartite system is named separable if it can be represented as a mixture convex combination of product states relative to the various parts of the total system, otherwise, it is termed entangled. Entanglement is considered a key ingredient in the increased efficiency of quantum computing compared to classical computation for certain quantum algorithms and thus it plays a central role in quantum information and communication.\cite{nielsenchuang,horodecki2009RMP}

However, entanglement does not exhaust the realm of quantum correlations. A quantum state of a composed system may contain other types of nonclassical correlation even if it is separable.\cite{Zurek2001PRL} In the framework of quantum-classical dichotomy, the total correlations present in the system can be separated in a purely quantum part and a classical part.  A measure of quantum correlations is the quantum discord.\cite{Zurek2001PRL,Henderson2001JPA} A mixed bipartite separable state with nonzero quantum discord may yet be exploited in quantum computation protocols.\cite{Knill1998PRL,Datta2008PRL,Lanyon2008PRL} An intense research activity has been dedicated
to the characterization of quantum discord for several classes of both bipartite\cite{Luo2008PRA,Ali2010PRA,Ferraro2010PRA} and multipartite\cite{Modi2010PRL} quantum
states, even in the case of continuous variable systems\cite{Adesso2010PRL,Giorda2010PRL}. An experimental observation of the quantum discord in nuclear magnetic resonance quadrupolar system was recently reported.\cite{Soares-Pinto2010PRA}

The dynamics of entanglement for bipartite quantum systems interacting with independent or common environments, either Markovian or non-Markovian, with the appearance of phenomena like sudden death\cite{yu2004PRL,yu2009Science}, revivals\cite{bellomo2007PRL,bellomo2008PRA} or trapping\cite{bellomo2008trapping}, can be considered well understood in its general lines. Dynamics of quantum discord has also received much attention for two-qubit systems in the presence of both Markovian\cite{werlang2009PRA,Ferraro2010PRA} and non-Markovian\cite{maziero2010PRA,fanchini2010PRA,wang2010PRA} environments. Interestingly, differently from entanglement, Markovian evolution can never lead to a sudden death of discord.
Quantum discord does result in fact to be more robust than the entanglement against decoherence. In some cases, in front of a decay of total correlations, classical and quantum correlations may present finite time intervals when each of them remains constant.\cite{mazzola2010PRL}

Recently, the transfer of entanglement between the parties of a composed system has been analyzed in several investigations.\cite{Cubitt2005PRA,lopez2008PRL,bai2009PRA} When two initially correlated atoms are placed in
two noninteracting leaky cavities, each connected to its own reservoir, it has been found that entanglement can be transferred from atoms to reservoirs via the cavities.\cite{Lopez2010PRA} In this case, different regimes have been shown to exist and in particular, under certain conditions, the cavities do not become entangled during the dynamics.
The reverse problem of entanglement transfer from radiation modes to qubits through cavities has been also investigated, finding the conditions for exchange of quantum correlations among the subsystems.\cite{Casagrande2007PRA,Bina2010EPL}
Here, we investigate the conditions for transfer of quantum correlations from qubits inside cavities to outside systems. In view of exploiting quantum discord, in this paper we analyze how it can be extracted for this kind of systems. The characteristics of discord transfer shall be compared with the ones known for entanglement transfer.

\section{Model}\label{par:model}

\begin{figure}
\begin{center}
\includegraphics[width=7.4 cm, height=4.5 cm]{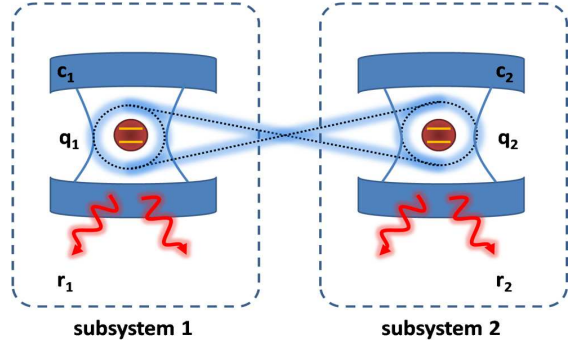}
\end{center}
\caption{\label{fig:system}\footnotesize (Color online) Schematic representation of the six-partite system. The two qubits $q_1$ and $q_2$ are initially entangled.}
\end{figure}
We consider a system composed by two noninteracting subsystems ($i=1,2$), each consisting in a qubit (two-level emitter) $q_i$ coupled to a single-mode cavity $c_i$ in turn interacting with an external reservoir $r_i$ (see Fig.~\ref{fig:system}). The Hamiltonian of the total system is thus given by the sum of the Hamiltonians of the two noninteracting subsystems
 \begin{eqnarray}\label{H totale}
     H_{tot} &=& H_1 + H_2.
\end{eqnarray}
In each subsystem, we distinguish the two-partite system of interest $S$ made by the qubit plus the cavity from the reservoir $r$ made by the external bosonic modes. The Hamiltonian of each part $i=1,2$ reads like (we omit index $i$, $\hbar=1$)
\begin{eqnarray}\label{H singola parte}
    H&=& H_S+H_r+H_I\, , \quad
     H_S =    \frac{1}{2} \omega_0 \sigma_z+ \omega_c\, a^\dag a + g (\sigma_-\, a^\dag  + \sigma_+\, a)\, , \nonumber \\
    H_r &=& \sum_k \omega_k b^\dag_k b_k,\quad
    H_I =\sum_k \kappa_k (a \, b^\dag_k + a^\dag \,b_k),
\end{eqnarray}
where $g$ is the coupling constant between qubit and cavity, $\kappa_k$ the coupling constants between cavity and external modes, $\sigma_z$ denotes the usual diagonal Pauli matrix, $\sigma_{\pm}$ are the two-level raising and lowering operators, $a$ and $b_k$ are the annihilation operators for the cavity and for the reservoir modes.

\subsection{Dynamics of subsystems}

Being the two subsystems noninteracting, they evolve independently so that we can analyze the dynamics of only one subsystem and use it to obtain the evolution of the global six-partite system. We will consider initial states such that, in each subsystem, only a single excitation is initially present in the qubit. This means that we will need to know the evolution of the single subsystem state $\ket{\varphi_0}=\ket{1}_q\ket{0}_c \ket{\mathbf{\bar{0}}}_r$, where $\ket{\mathbf{\bar{0}}}_r=\Pi_k \ket{0_k}_r$.
Under the action of the Hamiltonian of Eq.~(\ref{H singola parte}), $\ket{\varphi_0}$ evolves as
\begin{equation}
     \ket{\varphi_t}= \xi_t \ket{1}_q\ket{0}_c \ket{\mathbf{\bar{0}}}_r +\eta_t \ket{0}_q\ket{1}_c \ket{\mathbf{\bar{0}}}_r  + \Sigma_k \lambda_k  (t) \ket{0}_q\ket{0}_c \ket{1_k}_r ,
\end{equation}
which can be rewritten in terms of a collective state of the reservoir modes, $\ket{\mathbf{\bar{1}}}_r=(1/\chi_t)\Sigma_k \lambda_k (t) \ket{1_k}_r$, as
\begin{equation}\label{single part evolution}
    \ket{\varphi_t}=\xi_t \ket{1}_q\ket{0}_c \ket{\mathbf{\bar{0}}}_r +\eta_t \ket{0}_q\ket{1}_c \ket{\mathbf{\bar{0}}}_r + \chi_t\ket{0}_q\ket{0}_c \ket{\mathbf{\bar{1}}}_r.
\end{equation}
Eq.~(\ref{single part evolution}) allows to compute the joint evolution of the total six-partite system starting from an arbitrary initial state where only one excitation is present in each atom, as we will show explicitly in Sec.~\ref{sec:entanglement-evolution}. In the following we obtain the time dependent coefficients $\xi_t$, $\eta_t$ and $\chi_t$, which may in general be different for the two subsystems. For each subsystem $|\chi_t|^2=1-|\xi_t|^2-|\eta_t|^2$, so that it is sufficient to have $\xi_t$ and $\eta_t$. To this aim, we need to solve the reduced dynamics of the bipartite system of interest $S$ (qubit plus cavity) and compare its density matrix at time $t$ with the one obtained by tracing out the reservoir in the evolved state of Eq.~(\ref{single part evolution}), whose elements in the qubit-cavity basis $\mathcal{B}=\{\ket{1}\equiv\ket{11}, \ket{2}\equiv\ket{10}, \ket{3}\equiv\ket{01}, \ket{4}\equiv\ket{00}\}$ are
\begin{equation}\label{qubitcavitymatrix}
  {\rho_{S}}_{22} =|\xi_t|^2,  \; \;\; \; {\rho_{S}}_{33} =|\eta_t|^2,  \; \; \;\;{\rho_{S}}_{44} =|\chi_t|^2,   \; \;\;\; {\rho_{S}}_{23}={\rho_{S}}_{32}^*=\xi_t\eta_t^* .
\end{equation}
The reduced density matrix at time $t$ of the qubit-cavity system $S$, induced by the Hamiltonian of Eq.~(\ref{H singola parte}), may be obtained using a phenomenological master equation of the form
\begin{equation}\label{phenomenological master equation}
	\frac{\mathrm{d}}{\mathrm{d}t} \rho_S = i[\rho_S, H_S] + \frac{\gamma}{2} (2 a \rho_S a^\dag - a^\dag a \rho_S - \rho_S a^\dag a),
\end{equation}
where $\gamma $ represents the rate of loss of photons from the cavity. The above master equation is appropriate when the reservoir is at zero temperature, the coupling between the cavity and the external modes of the reservoir has a flat spectrum in the range of involved frequencies and the qubit is resonant with the cavity \cite{scala2007PRA,scala2007JPA}. We will limit our investigation to this physical condition, adopting therefore the master equation of Eq.~(\ref{phenomenological master equation}). The comparison between the solution of this master equation with the initial condition $\ket{1}_q\ket{0}_c $ and Eq.~(\ref{qubitcavitymatrix}) allows to obtain $\xi_t$ and $\eta_t$ as
\begin{equation}\label{xi and eta}
|\xi_t|^2 = \mathrm{e}^{-\frac{\gamma t}{2}}\left[\cos(\Omega t) + \frac{\gamma}{4 \Omega} \sin(\Omega t) \right]^2, \quad
|\eta_t|^2 = \frac{g^2}{\Omega^2 } \mathrm{e}^{-\frac{\gamma t}{2} }  \sin^2(\Omega t) ,
\end{equation}
where we have introduced the characteristic frequency $\Omega=\sqrt{g^2-\gamma^2/16}$. When $(4 g)/\gamma>1$ the functions $|\xi_t|$ and $|\eta_t|$ have a damped oscillatory behavior, while when $(4 g)/\gamma<1$ they become hyperbolic and oscillations disappear. When qubit and cavity are slightly out of resonance, the master equation of Eq.~(\ref{phenomenological master equation}) is expected to hold as well but the expressions of $\xi_t$ and $\eta_t$ become quite cumbersome.

In order to investigate the flow of classical and quantum correlations within the multipartite system, we first introduce some quantifiers able to distinguish quantum from classical correlations.

\section{Classical and quantum correlations}

As said before, entanglement does not necessarily exhaust all quantum correlations present in a state. An attempt to quantify all the nonclassical correlations present in a system, besides entanglement, has led to the introduction of the quantum discord, given by the difference between two expressions of mutual information extended
from classical to quantum systems\cite{Zurek2001PRL,Henderson2001JPA}. Following this framework, the total correlations between two quantum systems $A$ and $B$ are quantified by the quantum mutual information
\begin{equation}\label{Totale}
    \mathcal{I}(\rho_{AB})=S(\rho_{A})+S(\rho_{B})-S(\rho_{AB})
\end{equation}
where $S(\rho)=-\mathrm{Tr}(\rho \log_2 \rho)$ is the Von Neumann entropy and $\rho_{A(B)}= \mathrm{Tr}_{B(A)}(\rho_{AB})$.
It is largely accepted that quantum mutual information $\mathcal{I}(\rho_{AB})$ is the information-theoretic measure of
total correlations in a bipartite quantum state\cite{Luo2007PRA}.

On the other hand, the classical part of correlations is defined as the maximum information about one subsystem that can be obtained by performing a measurement on the other system. Given a set of projective (von Neumann) measurements described by a complete set of orthogonal projectors $\{\Pi_k\}$ and locally performed only on system $B$, the information about $A$ is the difference between the initial entropy of $A$ and the conditional entropy, that is $\mathcal{I}(\rho_{AB}|\{\Pi_k\})=S(\rho_{A})-\sum_k p_k S(\rho_k)$, where $\rho_k=(\mathbb{I}\otimes \Pi_k)\rho(\mathbb{I}\otimes \Pi_k)/\mathrm{Tr}[(\mathbb{I}\otimes\Pi_k)\rho(\mathbb{I}\otimes\Pi_k)]$, $p_k$ is the probability of the measurement outcome $k$ and $\mathbb{I}$ is the identity operator for subsystem $A$. To ensure that one captures all classical correlations, one needs to maximize $\mathcal{I}(\rho_{AB}|\{\Pi_k\})$ over all the sets $\{\Pi_k\}$. Classical correlations are thus quantified by $\mathcal{Q}(\rho_{AB})=\mathrm{sup}_{\{\Pi_k\}}\mathcal{I}(\rho_{AB}|\{\Pi_k\})$ and the quantum discord is then defined by
\begin{equation}\label{quantum discord}
    \mathcal{D}(\rho_{AB})=\mathcal{I}(\rho_{AB})- \mathcal{Q}(\rho_{AB}),
\end{equation}
which is zero only for states with classical correlations and nonzero for states with quantum correlations. The nonclassical correlations captured by the quantum discord may be present even in separable states.\cite{Zurek2001PRL}

The maximization procedure involved in computing the quantum discord has been analytically solved for certain class of quantum states. Along the paper, we will limit our treatment to a subclass of $X$-structured density operators which, in the standard two-qubit computational basis $\mathcal{B}=\{\ket{1}\equiv\ket{11},\ket{2}\equiv\ket{10},\ket{3}\equiv\ket{01},\ket{4}\equiv\ket{00}\}$, has density matrix elements given by
\begin{equation}\label{roX}
   \rho_{11}=a,\quad \rho_{22}= \rho_{33}=b, \quad\rho_{44}=d,\quad \rho_{14}=w,\quad \rho_{23}=z,
\end{equation}
where the coherences $\rho_{14}$ and $\rho_{23}$ are real numbers and $\rho_{22}=\rho_{33}$, so that $S(\rho_A)=S(\rho_B)$. Under this condition, the classical correlations assume the same value irrespective of whether the maximization procedure involves measurements performed on $A$ or on $B$.\cite{Zurek2001PRL,Henderson2001JPA} For this class of states, quantum discord is given by\cite{fanchini2010PRA}
\begin{eqnarray}\label{qd analitico}
    \mathcal{D}(\rho)&=&\mathrm{min}\{D_1,D_2\},\nonumber \\
    D_1&=&S(\rho_A)-S(\rho_{AB})-a \log_2\left(\frac{a}{a+b}\right)-b \log_2\left(\frac{b}{a+b}\right)\nonumber \\
    &&-d \log_2\left(\frac{d}{b+d}\right)-b \log_2\left(\frac{b}{d+b}\right),\nonumber \\
    D_2&=&S(\rho_A)-S(\rho_{AB})-\Delta_+ \log_2\Delta_+ -\Delta_+- \log_2\Delta_-,
\end{eqnarray}
with $\Delta_{\pm}=\frac{1}{2} (1\pm \Gamma)$ and $\Gamma^2=(a-d)^2+4(|z|+|w|)^2$.
Using this expression for quantum discord, classical correlations can be in turn obtained by Eqs.~(\ref{quantum discord}) and (\ref{Totale}).

In our analysis we will compare quantum discord and entanglement, quantified by concurrence\cite{wootters1998PRL} which varies from $C=0$ for a disentangled state to $C=1$ for a maximally entangled state. For states defined as in Eq.~(\ref{roX}), concurrence results to be $C=2\mathrm{max}\{0,z-\sqrt{a d},w-b\}$.\cite{fanchini2010PRA}

\section{Correlations transfer by input-output theory}\label{par:Transferofquantumcorrelations}

The dynamics of entanglement between two noninteracting qubits, each in its own cavity, has been analyzed by exploiting a model where each qubit is coupled to a continuous ensemble of cavity modes with a Lorentzian spectrum.\cite{bellomo2007PRL,lopez2008PRL} It is possible to show that this model is formally equivalent to the quasi-mode approach here adopted in which one distinguishes a single mode cavity in turn coupled to an external reservoir (see Appendix A). In particular, in the limit of continuous spectrum, the equivalence holds if the coupling constant $\kappa(\omega)$ between the cavity mode and a reservoir mode is independent of frequency over a band of frequencies about the characteristic cavity frequency $\omega_c$, that is $\kappa(\omega)\approx \kappa$. Using the input-output theory, an exact relation between the external modes and the intracavity mode may be obtained\cite{wallsbook}
\begin{equation}\label{eq:inputoutput}
	a_\mathrm{out}(t)+a_\mathrm{in}(t)=-i \sqrt{2 \pi}|\kappa|a(t),
\end{equation}
where the operators $a_\mathrm{out}$ and $a_\mathrm{in}$ are related to out-coming or incoming photons. In the case of no input photons, once known the quantum state for the cavity mode $a(t)$, it is possible to calculate quantities of interest for the output photons, like expectation values and correlation functions. These quantities can be either directly measured by photodetectors or used to know the amount of quantum correlations transmittable to other distant quantum systems. The approach here adopted allows us to follow how the two independent single-mode cavities mediate the flow of correlations from the atoms to the external environment and the building of quantum correlations between the cavities themselves. Following the above argument, these quantum correlations could be transferred to other systems by means of photons, escaping from each cavity, which may propagate into the free space or along an optical fibre until they eventually reach another distant bipartite quantum system.

\section{Dynamics of quantum discord and entanglement}\label{sec:entanglement-evolution}

In this section we investigate the dynamics of quantum discord and entanglement for some suitable couples of parties of the six-partite system. In particular, we will focus on the bipartite systems composed respectively by the two qubits, the two cavities and the two reservoirs. The qubits are initially in a two-excitation Bell-like state, while cavities and reservoirs are in their vacuum state. The two subsystems, 1 and 2, are considered identical for the sake of simplicity. The two-excitation initial state is
\begin{equation}
    \ket{\Psi_0}= \left(\alpha \ket{0_{q_1}0_{q_2}}+\beta \ket{1_{q_1}1_{q_2}}\right)\ket{0_{c_1}0_{c_2}}
    \ket{\mathbf{\bar{0}}_{r_1}\mathbf{\bar{0}}_{r_2}},
\end{equation}
with $\alpha$ real and $\beta$ complex. Using Eq.~(\ref{single part evolution}), the evolution of the total system is given by
\begin{equation}
   \ket{\Psi_t}=\alpha \ket{0_{q_1}0_{q_2}}\ket{0_{c_1}0_{c_2}} \ \ket{\mathbf{\bar{0}}_{r_1}\mathbf{\bar{0}}_{r_2}} + \beta  \ket{\varphi_t}_{1}
   \ket{\varphi_t}_{2},
\end{equation}
where the zero-excitation probability amplitude remains constant. From the state $\ket{\Psi_t}$ one finds the reduced density matrices of the bipartite system of interest tracing over the degrees of freedom of the noninvolved parties. In this way, e.g., the two-qubit state at time $t$ is given by
\begin{equation}\label{two-qubitmatrix}
 \hat{\rho}^\Psi_{q_1q_2}(t)=\left(
\begin{array}{cccc}
  |\beta|^2|\xi_t|^4 & 0 & 0 & \alpha\beta\xi_t^2  \\
  0 & |\beta|^2|\xi_t|^2(1-|\xi_t|^2) & 0 & 0 \\
  0 & 0 & |\beta|^2|\xi_t|^2(1-|\xi_t|^2) & 0 \\
  \alpha\beta^\ast(\xi_t^\ast)^2 & 0 & 0 & \alpha^2+|\beta|^2(1-|\xi_t|^2)^2 \\
\end{array}
\right).\\
\end{equation}
The concurrence corresponding to this density matrix is found to be\cite{bellomo2007PRL}
\begin{equation}\label{two-qubitconc}
C^\Psi_{q_1q_2}(t)=\mathrm{max}\left\{0,2|\beta||\xi_t|^2[\alpha-|\beta|(1-|\xi_t|^2)]\right\}.
\end{equation}
The two-cavity and two-reservoir density matrices at time $t$, $\hat{\rho}^\Psi_{c_1c_2}(t)$ and $\hat{\rho}^\Psi_{r_1r_2}(t)$, and their corresponding concurrences, $C_{c_1c_2}^\Psi(t)$ and $C_{r_1r_2}^\Psi(t)$, are easily obtained by Eqs.~(\ref{two-qubitmatrix}) and (\ref{two-qubitconc}) with the substitutions $\xi_t\rightarrow\eta_t$ and $\xi_t\rightarrow\chi_t$, respectively. Using Eqs.~(\ref{qd analitico}) and (\ref{two-qubitmatrix}), one can compute quantum discord between qubits, $D_{q_1q_2}^\Psi(t)$. With the substitutions $\xi_t\rightarrow\eta_t$ and $\xi_t\rightarrow\chi_t$, respectively, one obtains quantum discord between cavities, $D_{c_1c_2}^\Psi(t)$, and between reservoirs, $D_{r_1r_2}^\Psi(t)$. For all the three bipartite systems considered, classical correlations are equal to quantum discord, so that total correlations are just twice the quantum discord. Thus, it will be enough to consider the dynamics of quantum discord to obtain also the evolution of classical and total correlations.

\begin{figure}[h!]
\begin{center}
\includegraphics[width=5.75 cm, height=4.5 cm]{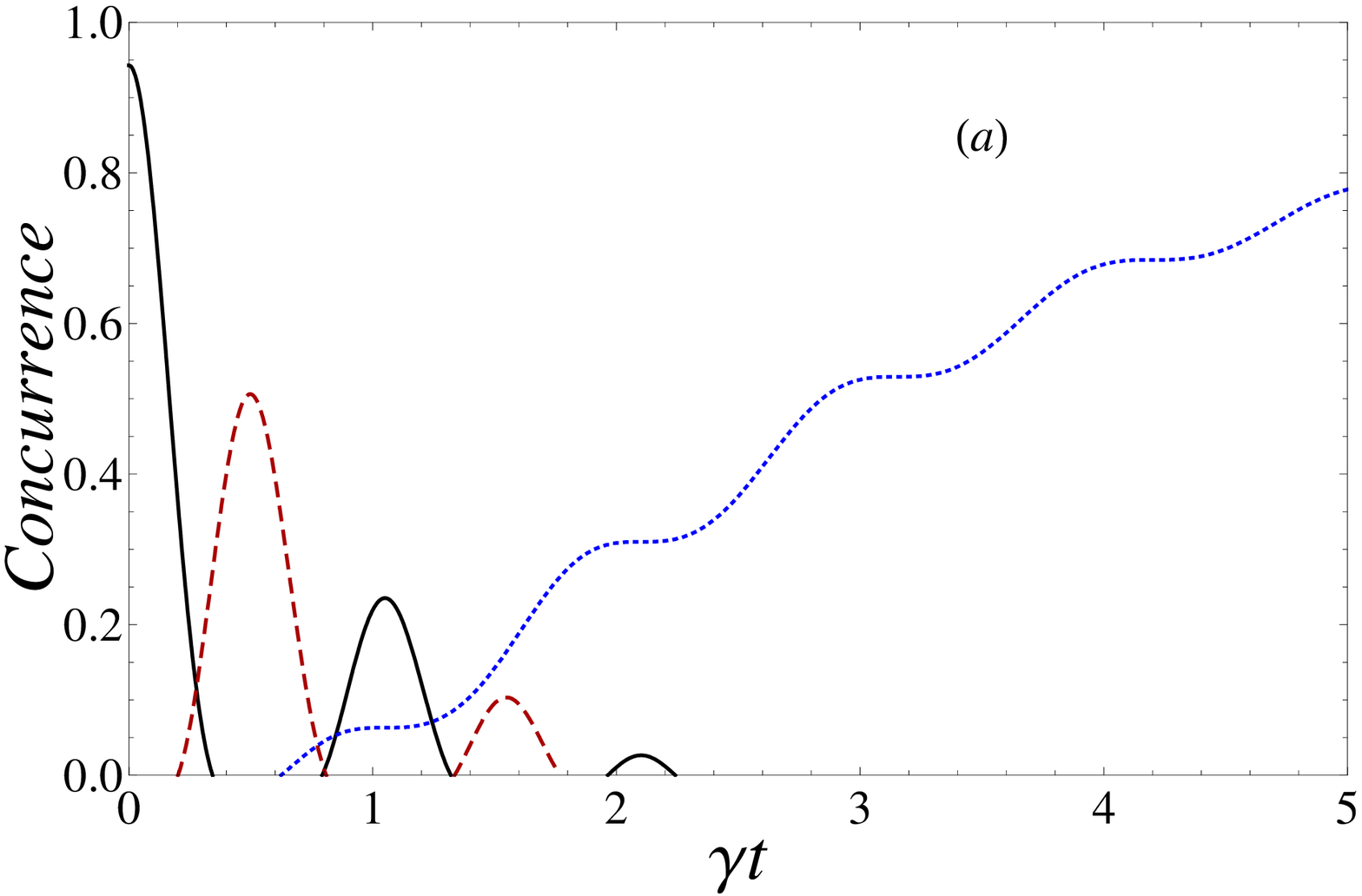}
\hspace{0.5 cm}
\includegraphics[width=5.75 cm, height=4.5 cm]{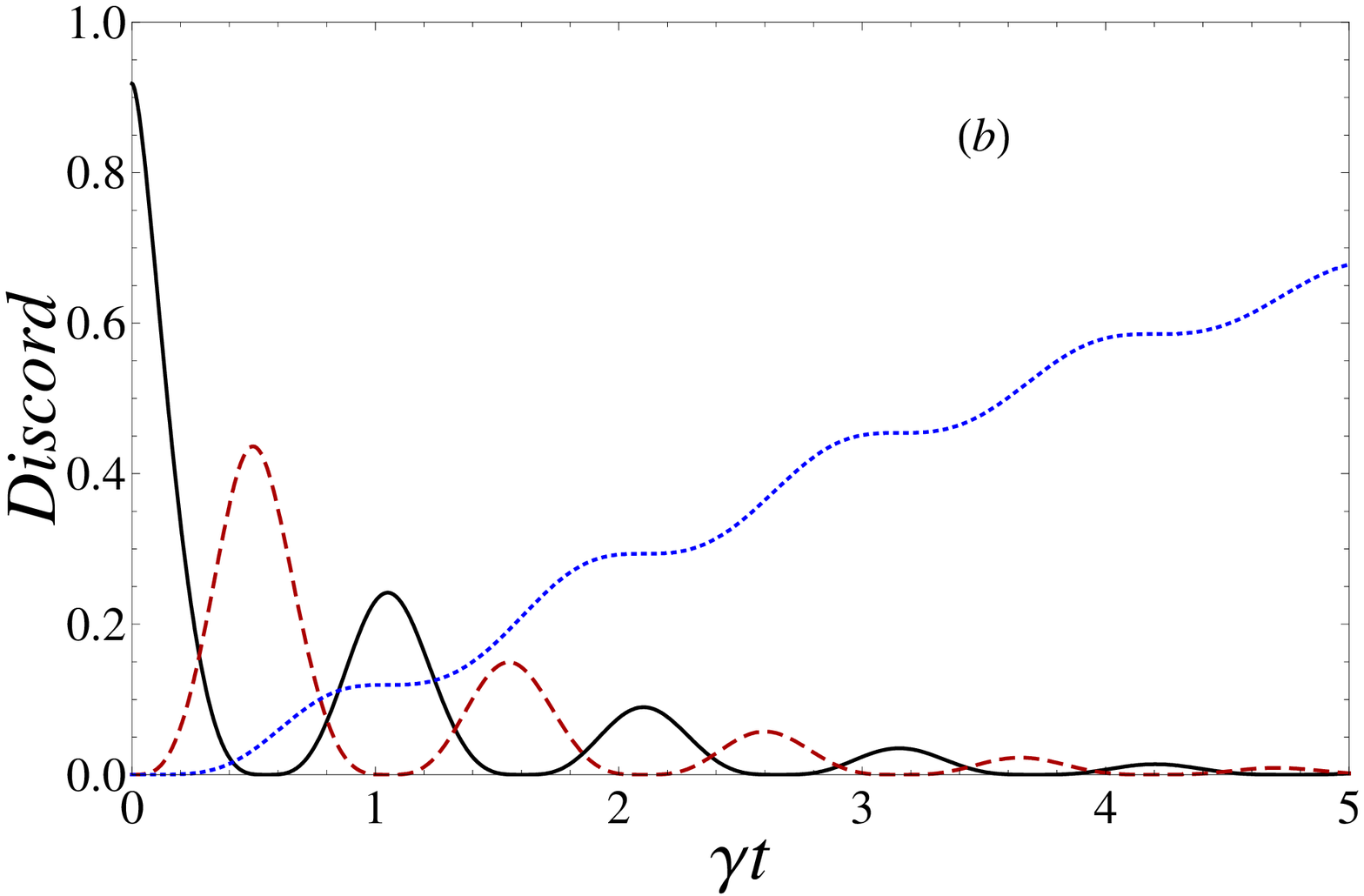}
\end{center}
\caption{\label{fig:concvsdisc}\footnotesize (Color online) Concurrence (panel a) and quantum discord (panel b)
as a function of the
dimensionless time $\gamma t$ for $\alpha=1/\sqrt{3}$ with $g=3\gamma$
for the two qubits $C^\Psi_{q_1q_2}(t), D^\Psi_{q_1q_2}(t) $ (black solid line), the two cavities $C^\Psi_{c_1c_2}(t), D^\Psi_{c_1c_2}(t)$ (red dashed line) and the two reservoirs $C^\Psi_{r_1r_2}(t), D^\Psi_{r_1r_2}(t)$ (blue dotted line).}
\end{figure}
In Fig.~\ref{fig:concvsdisc}, the three concurrences $C^\Psi_{q_1q_2}(t)$, $C^\Psi_{c_1c_2}(t)$ and $C^\Psi_{r_1r_2}(t)$ are plotted on panel (a) and the three quantum discords $D^\Psi_{q_1q_2}(t)$, $D^\Psi_{c_1c_2}(t)$ and $D^\Psi_{r_1r_2}(t)$ on panel (b) as a function of the dimensionless time $\gamma t$. From panel (\textit{a}) one sees that the initial entanglement between qubits is progressively lost with characteristic revivals.\cite{bellomo2007PRL} In front of this loss of correlations between qubits, the two cavities becomes entangled also showing entanglement revivals. After a certain time, correlations between the two reservoirs arise\cite{Lopez2010PRA}. From panel (b) one sees that quantum discord results to be more resistent than entanglement to decoherence effects. Discord between qubits progressively decays becoming zero at single instants and then rising again. In front of this process, the two cavities become quantum correlated already from the beginning with an oscillating decay behavior. Differently from what happens for entanglement dynamics, the reservoirs begin to present nonzero quantum discord already immediately after $t=0$.

\begin{figure}
\begin{center}
\includegraphics[width=5.75 cm, height=4.5 cm]{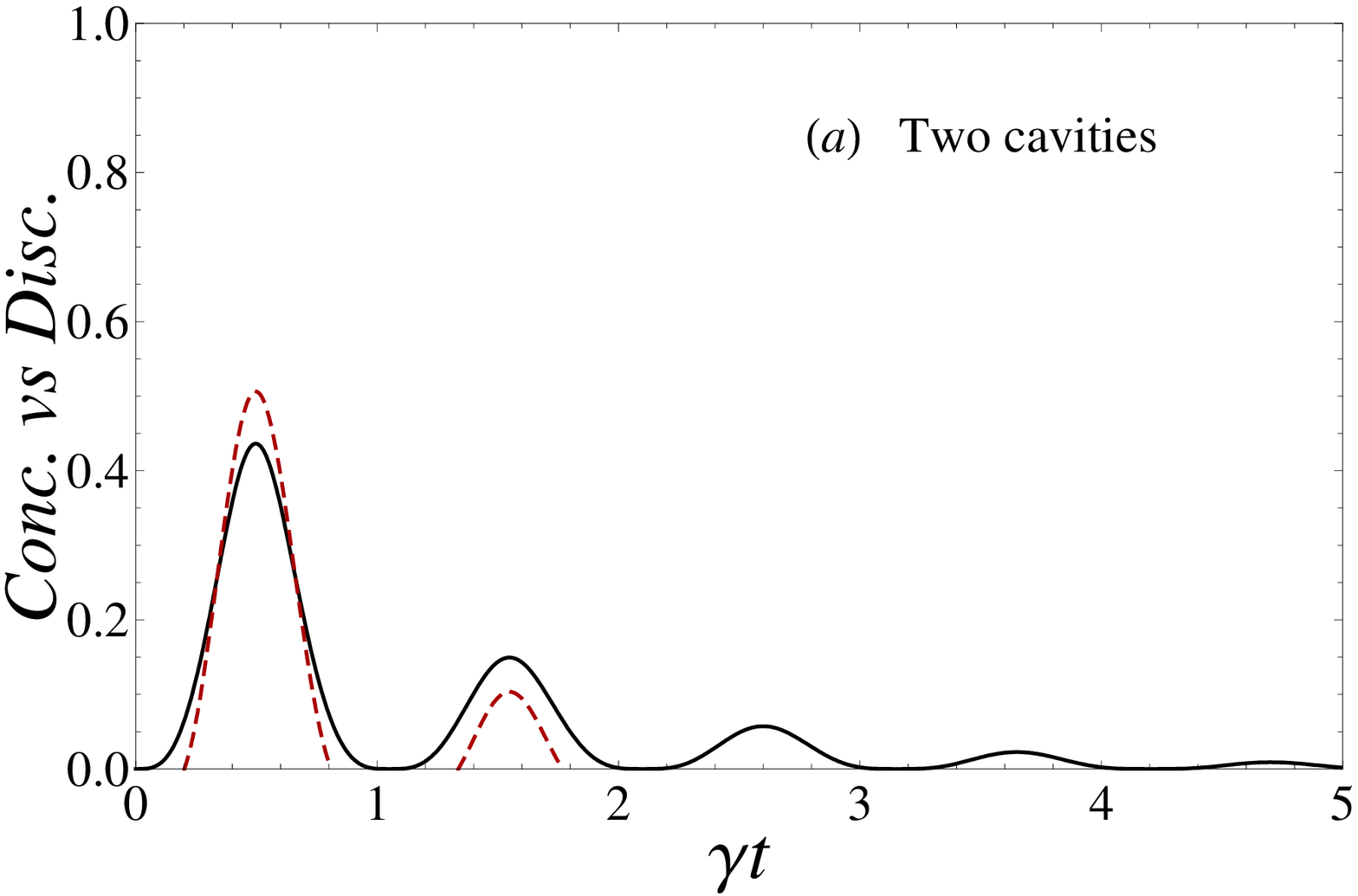}
\hspace{0.5 cm}
\includegraphics[width=5.75 cm, height=4.5 cm]{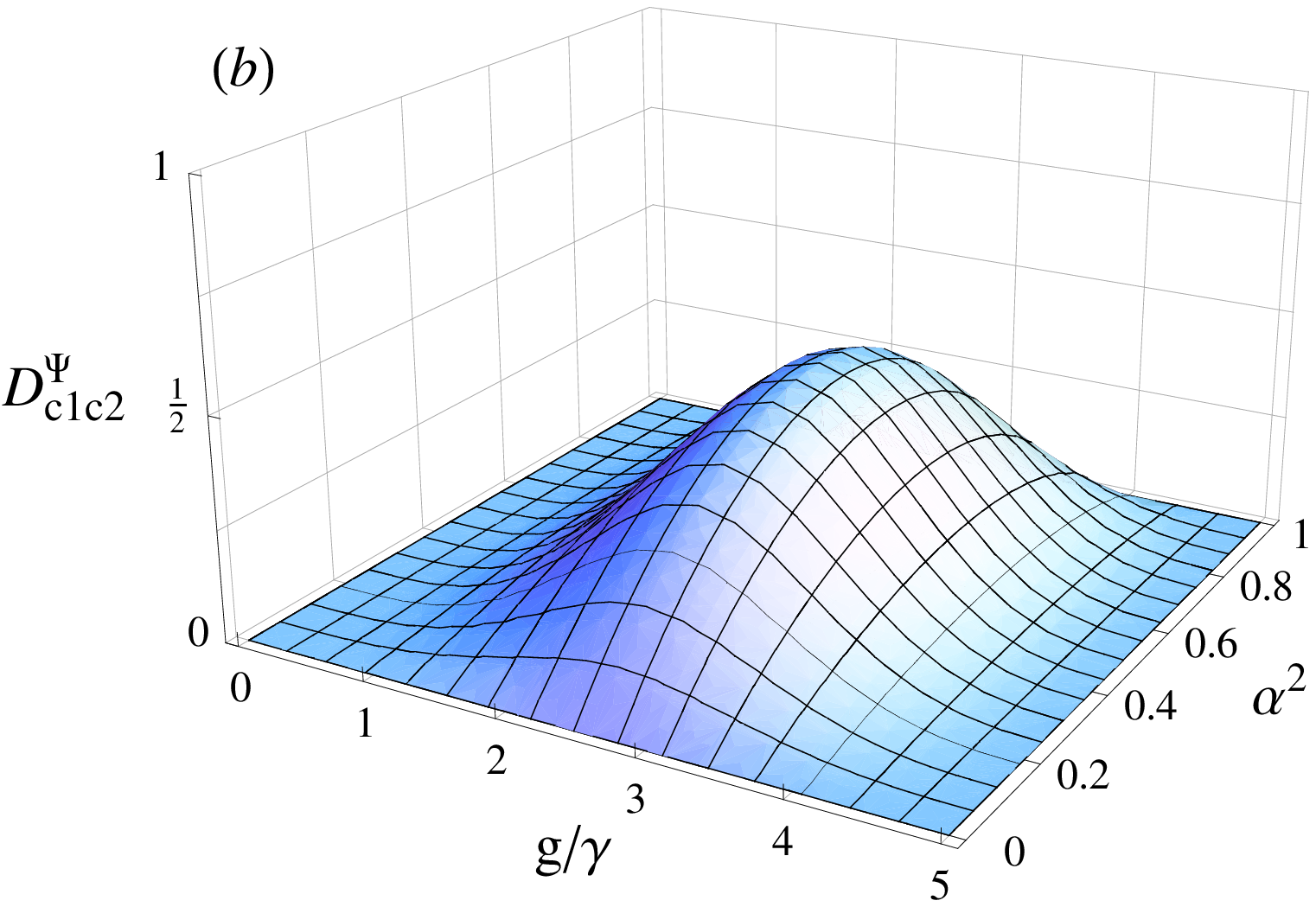}
\end{center}
\caption{\label{fig:entanglementvsdiscord}\footnotesize (Color online) Panel (a): concurrence (red dashed line) versus quantum discord (black solid line) for the two cavities as a function of the dimensionless time $\gamma t$ for $\alpha=1/\sqrt{3}$ with $g=3\gamma$. Panel (b): two-cavity discord, $D^\Psi_{c_1c_2}$ at a given time $\gamma t= 0.6$, as a function $g/\gamma$ and $\alpha$.}
\end{figure}
We are in particular interested in the dynamics of quantum correlations between the two cavities. In fact, the transfer of correlations to external systems is strictly connected to the correlations present between cavities (see Sec.\ref{par:Transferofquantumcorrelations}). In panel (a) of Fig.~\ref{fig:entanglementvsdiscord}, $C^\Psi_{c_1c_2}(t)$ and  $D^\Psi_{c_1c_2}(t)$ are compared. One sees that, with the exception of some time instants where quantum discord is zero, it can always be transferred from cavities to external distant systems, using the input-output theory of Sec.\ref{par:Transferofquantumcorrelations}. Differently, entanglement can be transferred only in limited time regions where the two-cavity state is not separable. It has been shown\cite{Lopez2010PRA} that, for small values of the parameters $\alpha$ and $g/\gamma$, entanglement can be transferred to reservoirs although the cavities do not get entangled during the process. Here, instead, the two cavities immediately become quantum correlated, the amount of correlations depending on the two ratios $\alpha/|\beta|$ and $g/\gamma$.
We also observe that for small times near the first peak of Fig.~\ref{fig:entanglementvsdiscord}(a), being concurrence a little larger than quantum discord, transfer of entanglement can be slightly more convenient than discord while for large times
the opposite occurs. In panel (b) of Fig.~\ref{fig:entanglementvsdiscord} the discord between cavities  $D^\Psi_{c_1c_2}$ is plotted at a given time $\gamma t=0.6$ as a function of both $g/\gamma$ and $\alpha$. The two-cavity quantum discord shows a non-monotone behavior in $g/\gamma$ and an almost symmetric dependence on $\alpha^2$.

The analysis presented here concerns the case when the two qubits are initially prepared in a pure entangled state. If the initial state of the qubits is mixed, one expects a similar behavior of concurrence and quantum discord for the various bipartite subsytems except that their initial values for the two qubits are smaller.

\section{Conclusion}\label{sec:conclusion}

In this paper we have investigated the transfer of quantum discord in a multipartite system consisting of two non interacting qubits each embedded in a single mode cavity in turn coupled to an external reservoir. The analysis of this model has allowed us to investigate how cavities mediate the flow of quantum correlations from qubit to the external environments. The comparison with the dynamics of entanglement has shown that quantum discord between qubits results to be more resistent than entanglement to decoherence effects. Differently from what happens for entanglement, cavities and reservoirs become from the beginning quantum correlated even when, in some time regions, their state is separable. In particular, the study of quantum discord between the two cavities show that, using the input-output theory, quantum correlations present in the cavities can be either transferred  to distant systems or externally measured at all times with the exception of some instants. This is not the case for entanglement which can be transferred only within finite time regions. In the last years the light-matter strong coupling regime has been achieved in single quantum dots-microcavity systems\cite{reithmaier2004Nature,yoshie2004Nature}. Very recently it has also been shown that ultracompact (with dimensions below the diffraction limit) hybrid structures, composed of metallic nanoparticles and a single quantum dot, can also achieve such strong coupling regime.\cite{savasta2010ACSNano} The investigation of the dynamics of quantum correlations here presented can also be applied to these solid state quantum devices.\cite{bellomo2011PhisScr}

\appendix
\section{The quasimode approach and Fano diagonalization}

We start considering  the following Hamiltonian describing the interaction between a single cavity mode with a continuum of bosonic modes,
\begin{eqnarray}\label{model}
    H_{\rm c}&=& \hbar \omega_{\rm a} a^{\dag}a + \int{d \omega \hbar \omega b^{\dag}\left(\omega\right)b\left(\omega\right)}
    + i \hbar \int{d \omega \kappa\left(\omega\right) a^{\dag}b\left(\omega\right)} + H.c.
\end{eqnarray}
This quadratic Hamiltonian can be diagonalized, following the Fano diagonalization method\cite{Fano1961PR,Savasta1996PRA}.
The eigenoperators of the system $c\left(\omega\right)$ that diagonalize the Hamiltonian can be written as a a linear combination of the cavity and bath modes, i.e.
\begin{equation}\label{diag2}
    c\left(\omega\right) = \alpha(\omega) a + \int{d\omega^{\prime} B\left(\omega,\omega^{\prime}\right)b\left(\omega^{\prime}\right)}\, .
\end{equation} where $\alpha(\omega)$ and $B\left(\omega,\omega^{\prime}\right)$ are coefficients to be determined.
The eigenoperators $c\left(\omega\right)$  have to satisfy the commutation relation $\left[c\left(\omega\right),H_{\rm c}\right]=\hbar \omega c\left(\omega\right)$. Inserting Eq.~(\ref{diag2}) into the previous commutation relation, the following system of equations is readily obtained
\begin{eqnarray}\label{sistema}
      (\omega_{\rm a}-\omega)\alpha(\omega) &=& i \int{d \omega^{\prime} B(\omega,\omega^{\prime}) \kappa^{*}(\omega^{\prime})}\\
         \omega \int{d \omega^{\prime} B(\omega,\omega^{\prime}) b(\omega^{\prime})}  &= & i \alpha(\omega) \int{d \omega^{\prime} \kappa(\omega^{\prime})b(\omega^{\prime})} + \int{d \omega^{\prime} B(\omega,\omega^{\prime}) \omega^{\prime} b(\omega^{\prime})}\, . \nonumber
\end{eqnarray}
System of Eq.~(\ref{sistema}) has peculiarities arising from its continuous spectrum. To solve it, we shall express $B(\omega,\omega^{\prime})$ in terms of $\alpha(\omega)$, utilizing the second equality of Eqs.~(\ref{sistema}), and enter the result in the first equality. From the second equality of Eqs.~(\ref{sistema}) we obtain
\begin{equation}
	 \omega B(\omega,\omega^{\prime})=\mathrm{i}\alpha(\omega)  \kappa(\omega^{\prime}) + B(\omega,\omega^{\prime}) \omega^{\prime}.
\end{equation}
This procedure now involves a division by $\omega-\omega^{\prime}$ which may be zero. Hence, the formal solution of the previous equation is
  \begin{equation}\label{bigrande}
    B(\omega,\omega^{\prime}) = \mathrm{i} \alpha(\omega)\kappa(\omega^{\prime})\left[{\cal P}\frac{1}{\omega -\omega^{\prime}}+ z(\omega)\delta(\omega -\omega^{\prime}) \right]\, ,
\end{equation}
where ${\cal P}$ indicates that the principal part is to be taken when the term is integrated and $z(\omega)$ will be determined in the following. Introducing Eq.~(\ref{bigrande}) in the first equality of Eqs.~(\ref{sistema}) we obtain
\begin{equation}\label{zeta}
	z(\omega)=\frac{\omega-\omega_a-F(\omega)}{|\kappa(\omega)|^2}\, , \quad \mathrm{where}\quad
F(\omega)={\cal P}\int{d \omega^{\prime} \frac{ |\kappa(\omega^{\prime})|^2}{\omega-\omega^{\prime}}}\ .
\end{equation}
In order to determine $\alpha(\omega)$, we impose the normalization condition for the $c(\omega)$, i.e. $[c(\omega),c^{\dag}(\omega^{\prime})]=\delta(\omega-\omega^{\prime})$, and obtain the following relation:
\begin{equation}\label{biglande}    \alpha(\omega)\alpha^{*}(\omega^{\prime})+\int{d\omega^{\prime\prime}B(\omega,\omega^{\prime\prime})B(\omega^{\prime},\omega^{\prime\prime})}= \delta(\omega-\omega^{\prime})\ .
\end{equation}
Introducing Eq.~(\ref{bigrande}) in Eq.~(\ref{biglande}), after some algebra, we obtain
\begin{eqnarray}
   |\alpha(\omega)|^2 &=& \frac{1}{|\kappa(\omega)|^2(z^2(\omega)+\pi^2)}=\frac{|\kappa(\omega)|^2}{(\omega-\omega_{\rm a}-F(\omega))^2+\pi^2|\kappa(\omega)|^2}\ .
\label{biglandissima}\end{eqnarray}
We notice that for a coupling $\kappa(\omega)$ independent on frequency, $F(\omega) = 0$.
It is possible to construct the operator $a$ in terms of the dressed operators $c(\omega)$. Using the commutator of $a$ with $c^\dag(\omega)$, $[a, c^\dag(\omega)]=\alpha^*(\omega)$, and inserting the expansion $a=\int d\omega\chi(\omega)c(\omega)$ into this commutator, we obtain $\chi(\omega)=\alpha^*(\omega)$.
The Hamiltonian term describing the interaction between the cavity mode $a$ and the single qubit can be now expressed as
\begin{equation}
	H_I = g \sigma_+\, a + H.c. = \sigma_+ \int d \omega \tilde g \alpha^*(\omega) c(\omega)\, .
\end{equation}



\end{document}